# UNDERSTANDING THE FOCUSING OF CHARGED PARTICLE FOR 2D SHEET BEAM IN A CUSPED MAGNETIC FIELD

Tusharika S Banerjee [1*], Arti Hadap, Reserach [2], K.T.V. Reddy [3]

[1]Scholar, Department of Electronics and Telecommunication Engineering, Sardar Patel Institute of Technology, Andheri, Mumbai-58

[2]Assistant Professor, Department of Physics, Mukesh Patel school of Technology, Management and Engineering, Vile-Parle, Mumbai-56.

[3]Director, PSIT, Kanpur, Agra, Delhi National Highway-2, Bhauti-Kanpur-209305

The requirement of axial magnetic field for focusing and transportation of sheet beam using cusped magnets is less as compared to solenoid magnetic fields which is uniform. There is often some confusion about how a cusped magnetic field focuses high current density sheet beam because it is generally understood that non-uniform magnetic field cannot guide the particle beam along its axis of propagation .In this paper, we perform simple analysis of the dynamics of sheet beam in a cusped magnetic field with single electron model and emphasize an intuitive understanding of interesting features (as beam geometry, positioning of permanent magnets, particle radius, particle velocity, radius of curvature of particle inside cusped magnetic field)

**Keywords**-Cusped magnetic field, Sheet beam, Space-charge, Diocotron Instability, Particle motion

## I. INTRODUCTION

Cusped magnetic field is an efficient option to focus charged particle sheet beam. Sheet beam carry high current density charged particles [1] and are used in high energy section of accelerators and other devices like BWO [2,3], TWT [4,5] and FELs [6,7] where production of high power microwaves is done [8,9]. The instability of sheet beam is the major constrain . The use of uniform magnetic fields [10,3], non uniform or cusped magnetic field [11,12], tailoring of beam edges [13,14] and variation in plasma density [15] have been theoretically predicted and experimentally demonstrated to overcome this limitation and it has been concluded that the requirement of axial magnetic field for focusing and transportation of sheet beam using cusped magnets is less as compared to solenoid magnetic fields which is uniform. [16,17] The use of cusped geometry for the generation of periodic fields (to achieve compactness) has stability [18] as its most noted advantage and is the most commonly used approach for generation of axial magnetic field( to achieve improved efficiency) in case of sheet beam. The possible disadvantage of cusped geometry is the uneconomical use of magnetic field and loss in different forms of energy. There has been different kinds of analysis on cusped magnetic fields like losses from leak width [19] and collection width [20], its effect on r.f. wave propogation [21], its use for ion confinement [22] and various aspects during use of cusped magnetic field like device construction [23], structure of magnetic field [24], guidance of divertor channels by cusped magnetic field [25], arrangements like diverging cusped field [26], efficient energy transfer [27], optimization of coil diameter [28] that generate cusped magnetic field, enhancement of plasma confinement with cusped magnetic field [29], cluster formation observation [30] etc. are few significant analysis that have been done for cusped magnetic field. However, the analysis on charged particles is limited when it comes to the efficient use of magnetic field using cusped magnetic field. Results on single particle trajectory have shown that diverging magnetic field of appropriate geometry and not the location of reflection point will alter the nature of trajectory. [31] It is essential to analyze group of particles or particle beam (2D), when the particle move like beams under the influence of cusped magnetic field. The analysis for cylindrical beam which are the most common, simple and easy to generate beams have been done .[32] The present paper emphasizes the analysis and study on the motion of group of particles inside sheet beam in detail for a cusped magnetic field. The present paper analyses the geometry of cusped magnetic magnetic field in terms of original thickness and changed thickness(due to space charge) of sheet beam . In Sec.II complete analysis of cusped magnetic field is done. With the developed equation of cusped magnetic field, the minimum permissible thickness of beam can be found. The allowed beam spread has been shown there. The particle radius is related to sheet beam thickness using single electron model.The exact positioning of magnets for the cusped arrangements is stated using dynamics of sheet beam.The behavior of important parameters that are related to the motion of charged particles like particle radius and particle velocity in case of sheet beam are discussed.Important results have come out upon discussions on beam spread, minimum thickness under given set of operating conditions ,positioning of magnets,the maximum and minimum value of space charge length, axial magnetic field ,the guiding characteristic based on particle radius and particle velocity in case of sheet beam . Conclusions are presented in Sec.III of the paper.

## II. ANALYSIS OF CUSPED MAGNETIC FIELD

The field in case of cusped magnetic field is non-uniform. Cusped magnetic field is used to improve efficiency.This kind of magnetic field not only reduces bulkiness but also reduces the input requirements considerably if permanent magnets are chosen as cusped arrangements for microwave generation.[33] It is important to state assumptions before discussions on the dynamics of a charged particle beam in cusped magnetic field. The space charge force is considered.The beam is 2D sheet beam and there is uniform distribution of particles all over.The beam is cold beam with initial transverse velocity as zero. Practically, both these assumptions are violated.The transverse velocity of particles inside beam leads to random distribution inside the beam.

### A. Equation of cusped Magnetic field

As the cusped magnetic field is applied to sheet beam, the beam gets unstable at space-change length .All the particles are assumed to have an initial velocity $v_z$ along the axis of propogation.The components of the axis symmetry magnetic field is given by

$$B_z(r,z) = B(z) \frac{-r^2}{4} B''(z) + \ldots \ldots (1)$$

$$B_r(r,z) = \frac{-r}{2} B'(z) + \frac{r^3}{16} B'''(z) + \ldots (2)$$

where z is the distance along the axis of beam prorogation , r is the radial distance from the axis of magnetic field.Under the paraxial approximation, terms up to the first order in r are taken in to consideration .It is assumed that $B(z)=B_0$ for $0<z<L$ and $B(z)=0$ otherwise.$B(z)$ is assumed to drop to zero at the ends. The analysis is performed in situation when beam gets dispersed or unfocused at space charge length L(RegionI) and comes back to its original state(Region II).This entire situation takes place along z,which is the axis of propogation of sheet beam.Region I and region II are shown in fig 1.We define two regions corresponding to z: z<0<L(Region I ) and z>L(Region II). The original sheet beam with thickness t is at emission point E.At space-charge length L, which is point F, the unfocused beam with maximum dispersed thickness denoted by t'(under given set of operating conditions) is shown. It is assumed that at point G,the beam retains its original thickness t.We will assume the maximum value of L as the maximum or final changed value of thickness denoted by t' is observed at beam spread. We will show later through the beam dynamics that the maximum value of L would be t'.

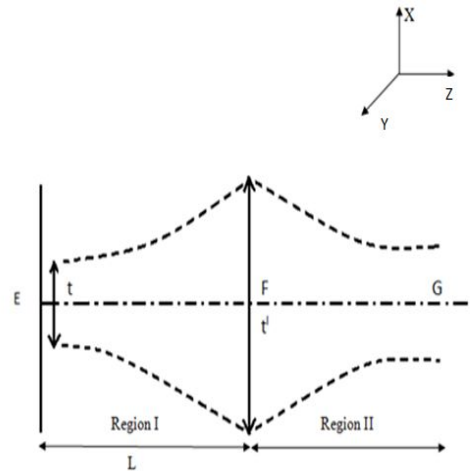

Fig.1: Typical waveform of cusped magnetic field

In case of paraxial approximation, expression for magnetic field can be written as

$$B_z = B_0[u(z) + u(-z+L)] \qquad (3)$$

$$B_r = \frac{-r}{2}B_0[\delta(z) - \delta(z-L)] \qquad (4)$$

where $t/2 < u(z) < t'/2$ for $z > 0$ and $\delta(z)$ is Dirac delta function here. The magnetic field can be written as summation of two functions.

$$B_z = B_0[X_1(z) + X_2(z)] \qquad (5)$$

and $X_1(z)$ and $X_2(z)$ are the fields in region I and region II respectively shown in fig 2.

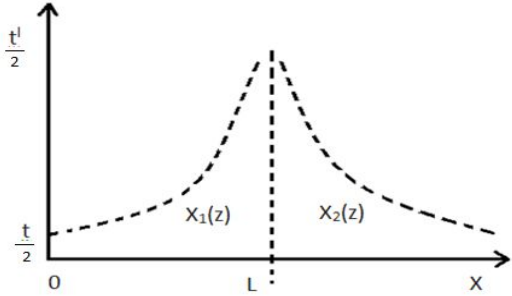

Fig:2:Two sided cusped magnetic field on the upper side of propogation axis

It has been observed and analysed [34] that while the power law predicts the field free region more accurately, the exponential nature will describe the field regions. The overall equation of axial cusped magnetic field on the upper side of propogation axis, when applied to a sheet beam with changed thickness t' is given by

$$B_z = B_0 \frac{t'}{t'-2z}[1 + (\frac{2z}{t'})^L] \qquad (6)$$

The axial magnetic field is related to the maximum changed thickness t' of the sheet beam. It is dependent on space charge length L along the axis of propogation (z here). The requirement of axial magnetic field is infinity if the positioning of magnets are at maximum changed value of t(L=t') or if the beam is allowed to spread as much as possible under given set of operating conditions.

The above expression gives a clear understanding on permissible beam spread or allowed change in thickness of sheet beam. It is evident that at 25% of $L_{max}$, the beam requires thrice axial magnetic field than at 50 % value of $L_{max}$. However, very small change in the value of axial magnetic field is observed at 75% value of $L_{max}$ as required at 50% of $L_{max}$. This indicates the positioning of magnets should be somewhere between 50% to 75% value of $L_{max}$. The requirement of axial magnetic field is infinity at 100% value of Lmax. Also, the positioning of magnet should be avoided beyond 90% of $L_{max}$, else large requirements of magnetic field will be there. It is true that the magnets should be placed as far as possible to avoid bulkiness but the beam quality will be compromised with positioning of magnet beyond 50% as evident from the above expression. Since, the reduction in axial magnetic field by three times is observed at 50%, for optimum outputs, 50% of $L_{max}$ can be chosen as the position where magnets should be placed. However the choice of length can go upto 80% if bulkiness is the major criteria.

**B. Analysis on sheet beam geometry (based on beam dynamics ) in cusped magnetic field**

The particle behavior points three classes of charged particles in cusped magnetic field. The first being very stable that move back and forth far away from null point and radiate energy over the time, the second type that move to center point with no magnetic field having infinite gyroradius whose motion is straight line and the third type of particles are a transition between the two. The non-adiabatic motion of a single particle in case of cusped magnetic field has been analyzed [34]. The electron off centering nature arises from non zero radial component of particle velocity on the upstream side of the cusp transition and another arises from finite width of cusp transition [35]. In cusped magnetic field, longitudinal magnetic field is peaked at the point cusp and approaches zero at the line cusp. In contrast, the radial magnetic field is maximum at the line cusps and minimum at point cusp.

**a. Particle radius inside cusped magnetic field**

In Fig. 3, the periphery of the cross sectional beam is shown by solid line as it enters the field, that is, region II. Eight particles are located at 1,2,3,4,5,6,7 and 8.

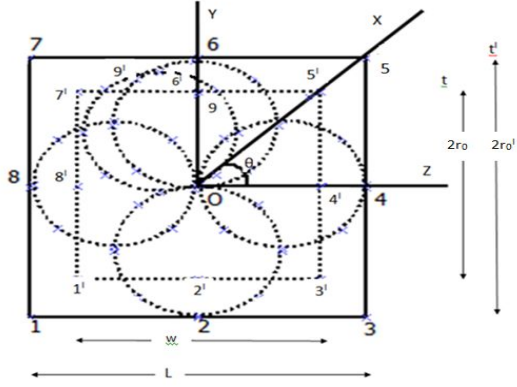

Fig. 3: Pictorial explanation of focusing of charged particle in 2D sheet beam. The solid boundary shows the electron as it enters the field. The dashed boundary shows the periphery of the electron beam after it travels some distance in field. The small circles are the trajectories of individual electrons.

The changed beam thickness is t'.The value of t' is 2r$_0$'. After the beam travels a certain distance through field, these particles move, from location 1,2,3,4,5,6,7 and 8 to 1',2',3',4',5',6',7' and 8'..Particle located at 9 of individual electron trajectory is initially inside the periphery of beam which corresponds to the stable.

L is the distance at which field lines are supplied.r$_0$ is the particle radius and is equal to half sheet beam thickness.

$$r_0 = \frac{t}{2} \quad (6)$$

Similarly, r$_0$' is half the value of dispersed thickness of sheet beam..r$_0$' is the radial co-ordinate of 6 after beam has dispersed due to space charge effect.

$$r'_0 = \frac{t'}{2} \quad (7)$$

The change in particle radius inside region II is given by following

$$\Delta r_0 = \frac{t'}{2} - \frac{t}{2} \quad (8)$$

The change in particle radius is 0.5 times of the changes value of the thickness of 2D sheet beam.

**b.Space charge length L**

Let us consider the right angled triangle 045 (from Fig.3)and by Pythagoras theorem the value of changed sheet beam thickness t' is given by

$$\frac{t'}{2} = \frac{L}{2}\sec\theta \quad (9)$$

where L is the propogation length with significant space charge effect of expanded sheet beam thickness t'.L is the space charge length.Thus, the value of following length L is given by

$$L = t'\cos\theta \quad (10)$$

where θ is the angle made by electron beam with the axis of propogation. The maximum and minimum value of space charge length,for given set of operation,that use sheet beam can be found as per the above developed relation.It can be concluded that with electrons of the beam being perfectly aligned with propogation axis, maximum value of L is possible.

$$L_{max} = t' \quad (11)$$

It is interesting to note from above relation that maximum value of L will be equal to changed thickness t'.The minimum value of L will be 0. Consequently,it means immediate positioning of magnets at the emission point of charge particles.

**c.Analysis on radius of curvature of particle inside cusped magnetic field**

The individual trajectories of electrons will be rising and decaying exponential functions as shown in fig.2. In case of solenoidal or uniform magnetic field,the radius of curvature or particle radius for cusped magnetic field is given by

$$R_c = \left|\frac{\gamma_m v_\theta}{eB_0}\right| = \frac{r_0}{2} \quad (12)$$

In the above expression, m is the electron's rest mass, e is the magnitude of electronic charge, and $\gamma = \sqrt{1 - \frac{v_z^2}{c^2}}$ is usual Lorentz factor. $r_0$ is the radial co-ordinate of the particle. $r_0$ is commonly known as the particle radius. This particle radius has different values at original and dispersed sheet beam thickness. The particle radius in case of original sheet beam (Instant1) is $r_0 = t/2$ and from (6), the above equation can be expanded further as

$$R_c = \left|\frac{\gamma_m v_\theta}{eB_0}\right| = \frac{r_0}{2} = \frac{t}{4} \quad (13)$$

Similarly, at the dispersed thickness t' of sheet beam $r_0 = t'/2$ from (7). It can be interpreted from (11) that the radius of curvature of particle in cusped magnetic field is one fourth of sheet beam thickness.

In case of cusped magnetic field with sheet beam at any instant, the radius of curvature inside region II for every off-axis particle is one fourth of the changed thickness at that instant. This is the radial displacement of particle from axis of cusped magnetic field. Let us discuss this.

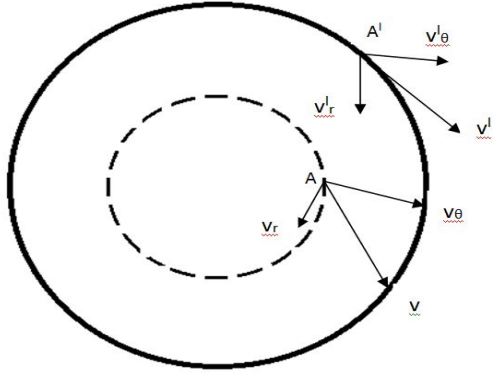

Fig.4: Particle velocity decomposition into radial and azimuthal components. A' is the waveguide point and A is the point with no magnetic field.

From Fig.4, at the anode point of waveguide where sheet beam thickness is t' the radius of curvature Rc' can be written as

$$R'_c = \left|\frac{\gamma_m v'_\theta}{eB_0}\right| = \frac{t'}{4} \quad (14)$$

Similarly at any arbitrary location in region II (between A and A' ), the radius of curvature will be given as

$$R''_c = \left|\frac{\gamma_m v''_\theta}{eB_0}\right| = \frac{t''}{4} \quad (15)$$

From Fig 4, radius of curvature Rc at A and A' that correspond to original and changed beam thickness is given by

$$\Delta R_c = \frac{t'}{4} - \frac{t''}{4} \quad (16)$$

where t'' is the sheet beam thickness at the arbitrary location of region II. The change in radius of curvature at any instant inside a cusped magnetic field is equal to one fourth of the change in thickness of sheet beam.

At 25% value of thickness, the radius of curvature of cusped magnetic field is observed. In order to maintain the original shape of sheet beam, the radius of curvature of particle inside cusped magnetic field must be equal to one fourth of sheet beam thickness.

**d. Particle velocity inside cusped magnetic field**

The angular velocity of the particle about the axis of magnetic field in cusped arrangement shall not remain constant. The radial component of the velocity will remain continuous when the particle enters region II from region I. So, $v_r = 0$ at z=0 in region II. Let us now analyze the angular velocity change in detail. The particles rotate with the cyclotron frequency around the center of their individual trajectories but rotate with the Larmour frequency around the axis. Larmour frequency is defined as $\omega_L = eB_0/2\gamma m$ and cyclotron frequency $\omega_c$ is defined as $\omega_c = eB_0/\gamma m$. The force in azimuthal direction is $-ev_z B_r(z)$ Due to $B_r(z)$, increment $\Delta v_\theta$ is given by,

$$\Delta v_\theta = r_0 \frac{eB_0}{2\gamma m}$$

As it crosses boundary. It must be noted from fig. 3 for point 6 and 6', along x axis, the change in $r_0$ will be $\Delta r_0$. Let us put the value of $r_0$ from (12) Consequently

$$\Delta v_\theta = \frac{eB_0}{\gamma m}[t'-t] \qquad (17)$$

The angular velocity increases as the beam spreading takes place. At the time of application of cusped magnetic field, the incremented value of $v_\theta$ can be seen which is defined by the above relation. No increase in $v_\theta$ is observed if sheet beam thickness does not change and more and more value of $v_\theta$ will imply more spread in sheet beam. The cusped magnetic field thus reduces this velocity in azimuth direction.

## B. Diocotron Instability-Analysis of linear shift Δd (based on single particle linear theory)

Sheet beam is large current density. If a close look on various kinds of instabilities are taken into consideration, apart from the instability caused due to space charge, the most dominant instability that will occur in sheet-beam is diocotron instability. This instability is created by two sheets of charge that slip past each other. It is shown in Fig.5. In simple words it is the rectilinear shift from the path of propogation. We next analyze the diocotron instability

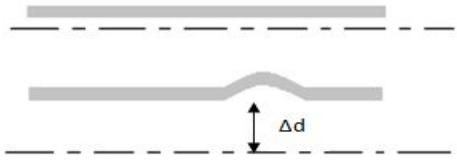

Fig. 5 : The rectilinear shift of sheet beam along it's axis of propogation due to linear displacement d.

The cross- section (CS), for sheet beam is taken elliptical since these elliptical cross-sections are easy to both focus and generate. [36,37]

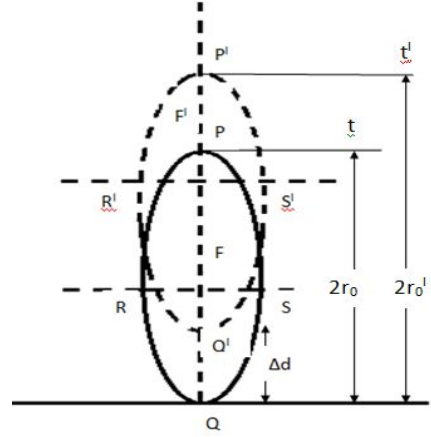

Fig. 6 : 2D sheet beam CS. Solid boundary is the periphery of beam CS with PQ as as major axis and RS a minor axis. Dotted boundary is the shifted periphery of beam.

Elliptical cross-sections are made by tailoring of the beams around the edges . It is one of the ways to control the sheet beam instability.

In the above figure, PQ is the CS of sheet beam and P'Q' is the shifted CS of sheet beam. The shift in focus F to F' is seen. There is a linear shift of the elliptical CS beam, which is vertical to the propogation axis. It must be observed that the original thickness PQ is same as the shifted diameter P'Q'. The the original sheet beam thickness t is to be achieved after focusing. Also, the expanded beam with thickness t' is P'Q. It can be said that the linear displacement $\Delta d$ from its original path will be given by

$$\Delta d = 2 \Delta r_0 \qquad (18)$$

From fig.6, $2r_0$= P'Q and $2r_0'$= P'Q. From equation (8)

$$\Delta d = t' - t \qquad (19)$$

It is evident from above relation that at higher values of input current, t' increases and so does the diocotron instability. The value of thickness has to be kept high to match the value of t' as per above relation. Especially in cases where there is more instability or higher values of t' are observed, thicker beams that match t' should be taken. However, w x t ratio has to be maintained. This limits the choice to take any large value of t. As usual the current density is compromised too. Exact magnitude of diocotron instability can be calculated from (19).
From (11), the above relation can be written as

$$\Delta d = L_{max} - t \quad (20)$$

The diocotron instability can completely be stabilized by increasing the beam thickness to the maximum value of space charge length.The maximum value of L practically if chosen to be 50% of t' for optimum outputs,

$$\Delta d = \frac{t'}{2} - t \quad (21)$$

The minimum chosen thickness could be half the changed thickness of sheet beam under a given set of operating conditions in order to overcome the diocotron instability which is most prominent in sheet beam.

## IV. CONCLUSION

The cusped magnetic field helps to achieve compactness.A single magnet is broken into N small magnets, thereby observing the weight reduction (between 1/N to $1/N^2$ ).The use of N magnets improves efficiency.The guiding characteristic of cusped magnetic filed in case of high current density sheet beam is because of two major reasons, as the behavior of charged particles inside sheet beam are analyzed. The first being the fact that longitudinal component of axial magnetic field for cusped magnets is maximum at point cusp and minimum at line cusp.Secondly,while the radial velocity does not change and is continuous, there is significant increase in azimuthal velocity.The cusped magnetic field reduces the azimuthal velocity so that beam is stable.In case of cusped magnetic field for sheet beam while the particle radius is half of the sheet beam thickness, the radius of curvature of particles in sheet beam in cusped magnetic field is one fourth of sheet beam thickness.The requirement of axial magnetic field is maximum at the position where the beam spread is equal to space charge length.The minimum chosen thickness, (in most cases generally) must be half of the changed thickness of sheet beam to completely overcome diocotron instability.The maximum change in thickness of sheet beam should not be allowed else requirement of infinite axial magnetic filed in case of cusped magnets shall take place. Practically, it is very difficult to discipline the sheet beam at this situation.However, the magnets should be placed at 50% to 80% of maximum changed thickness. The beam spread should not be allowed beyond 50% to 80% of its maximum spread.

**References**

[1]Booske,J.H.,W.W. Destler, Z. Segalov, D. J.Radack,E.T.Rosenbury, .Rodgers,T.M.Antonsen,Jr.,V.L.Granaststein, and I.D. Mayergoyz ,"Propogation of wiggler focused relativistic sheet electron beams",Appl. Phys. 64, 6(1988).

[2]Banerjee,T.S., Arti Hadap and KTV Reddy, "Review on Microwave generation using Backward Wave Oscillator",Scholars Research Library Archives of Applied Science Research 04, 129–135. (2014).

[3]Wang,J.,Yubin Gong , Yanyu Wei, Zhaoyun Duan,Yabin Zhang, Linna Yue, Huarong Gong, Hairong Yin, Zhigang Lu, Jin Xu and Jinju n Feng," High power millimeter-wave BWO driven by sheet electron beam", IEEE transactions on Electron devices 60, 471-476(2013).

[4]Field,M., Robert Borwick,Vivek Mehrotra, Berinder Brar, J. Zhao,Young -Min Shin, Diana Gamzina, Alexander Spear, Anisullah Baig, Larry Barnett, Neville Luhmann, Takuji Kimura,John Atkinson,Thomas Grant ,Yehuda Goren and Dean E. Pershing, " 1.3:220 GHz 50 W Sheet beam traveling wave tube amplifier",IEEE transactions on Terahertz Science and Technology 10, 21-22(2010).

[5]Shin,Y.M., Anisullah Baig, Diana Gamzina and Neville C. Luhmann,Proceedings of IEEE International Vacuum Electronics Conference(IVEC),18-20 May 2010 ,Monterey, CA, " MEMS fabrication of 0.22 THZ sheet beam TWT circuit",(IEEE,2010) pp. 185-186.

[6]Cheng,S.,W.W.Destler,V.L. Granatstein, T.M. Antonsen,B. Levush, J.Rodgers and Z.X. Zhang" A high millimeter wave sheet beam free electron laser amplifier", IEEE transactions on Plasma Science24, 750-757(1996).

[7]T. C. Marshall, " Free Electron Lasers" (MacMillan, New York, 1985).

[8]Booske,J.H., Richard J. Dobbs,Colin D. Joye, Carol. Kory, George R. Neil, Gun-Sik Park, Jaehun Park and Richard J. Temkin."Vacuum electronic high power terahertz sources",IEEE transactions on Terahertz science and technology 1, 54-75(2011).

[9]Wang,J.,Yiman Wang,Lili Li, Yanchu Wang, Wei Liu, A. Srivastav, J.K. So and G.S.P, "Generation and application of high current density sheet beams for THz vacuum electron sources",Proceedings of 32nd International conference on Infrared and Millimeter waves and 15th International conference on Terahertz Electronics,2-9 Sept. 2007 Cardiff,(IEEE,2007) pp. 260-261.

[10]PandaP.C., Visnu Srivastava and Anil Vora "Analysis of Sheet Electron Beam Transport Under Uniform Magnetic Field",IEEE transactions on Plasma Science 41,461-469(2013).

[11]Basten,M.A.,J.H.Booske and J.Anderson, "Magnetic quadrupole formation of elliptical sheet electron beams for high-power microwave devices", IEEE transactions on Plasma sciences 22,960-966(1994).

[12]Booske,J.H. and Brian D. Mcvey,"Stability and confinement of non-relativistic sheet electron beams with periodic cusped magnetic focusing", J. Appl. Phys. 73,4140-4155 (1993).

[13]Booske,J.H.,M.A Basten,A.H. ,Kumbaskar, T.M.AntonsenJr.,S.W.Bidwell,Y.Carmel,W.W. Destler, V.L.Granatstein and D.J. Radack"Periodic magnetic focusing of sheet electron beam",Physics of Plasma 1, 1714-1720(1994).

[14]Panicker,J.,Y.Choyal,K.P.Maheshwari and Sharma, "Space-charge electrostaic fields and focusing of a sheet electron beam with diffused edges",Laser and particle beams 17,1-13. (1999).

[15]Gokhale,A.,Preeti Vyas,J Paniker,Y Choyal and KP Maheshwari"Numerical investigation of space charge electric field for a sheet electron beam between two conducting planes",Pramana 58, 67-77(2002).

[16]Carlsten,B.E.,"Technology development for a mm-Wave Sheet -Beam Travelling-Wave Tube",IEEE transactions on Plasma Science 33,85-93 (2005).

[17]Hadap,A. & K.C. Mittal" High Current Planar Beam in a Wiggler Magnet Array",IEEE Pulsed Power Plasma Science Conference, (2007).

[18]Berowitz,J., H Grad and H Rubin, in proceedings of the second United Nations International conference on peaceful uses of atomic energy, Geneva, 1958, Vol 31, Page 177 .

[19]Kozima H., S.Kawamoto and K. Yamagiwa. " On the leak width of line- and point-cusp magnetic fields ",Science direct Physics Letters A 86, 373-375,(1981).

[20]Hubble A. Aimee and John E. Foster, "Primary Electron Transport and Losses in a Ring-Cusp Discharge Chamber"Presented at the 31st International Electric Propulsion Conference,(American Institute of Aeronautics and Astronautics),pp.1-12 (2009).

[21]Takechi S., Shunjiro Shinohara and Yoshinobu Kawai, " Effects of cusp magnetic field configuration on wave propogation in large diameter r.f. produced plasma", Elsevier Sci. Surface and Coating Tech.112, 15-19,(1999).

[22]Hershkowitz N., John R and Hideo Kozima, "Electrostatic self-plugging of a picket fence cusped magnetic field",Phys. Fluids 22, 122-125 (1979).

[23]Combes L.S., C.C. Gallagher & M.A. Levine, "Plasma Behavior in a Cusp-Mirror Magnetic Field",Phys. Fluids 5, 1070 (1962).

[24]Dankongkakul B. , Samuel J. Araki & richard E. Wirz, " Magnetic field structure influence on primary cusp looses for micro-scale discharges", Phys. of plasmas, 21, 043506-1-043506-7 (2014).

[25]Takase Haruhiko, "Guidance of divertor channel by cusp like magnetic field for tokamak devices", J. Phys.LettersSoc. Jpn.,70, 609-612 (2001).

[26]Courtney Daniel G.& Manuel Martinez-Sianchez, "Diverging cusped field hall thruster", Presented at the 30th International Electric Propulsion Conference,Florence,Italy,17-20 Sept.pp.1-10 (2007).

[27]Lin C.H., P. W Chen and Ching Yao Chen , "Simulations of silicon cz growth in a cusp magnetic field",Magnetohydrodynamics 47, 17-28 (2011).

[28]Wang Y., Wenting Xu Xiaolin Dai, Qinghua Xiao, Shujun Deng Zhirui Yan & Qigang Zhou, " Numerical simulation of effects of cusp magnetic field on oxygen concentration of 300 MM CZ-Si",Springer link,Rare materials 31,494-499 (2012).

[29]Ark J., Nicholas A. K., Paul E. Sieck, Dustin T Offerman, Michael Skillicorn, Andrew Sanchez, Kevin Davis,Eric Alderson &Giovanni Lapenta, "High energy electron confinement in a magnetic cusp configuration",Phys.Rev.X 5,021024-1-021024-10 (2015).

[30]Nykyri K., P.J. Cargill, E. Lucek, T. Horbury, B. Lavraud, A. Balogh, M.W. Dunlop, Y. Bogdanova, A. Fazakerley , I. Dandouras & H. Reme, "Cluster observations of magnetic field fluctuations in the high-altitude cusp",Annales Geophysicae, European Geosciences Union 22 (7), 2413-2429(2004).

[31]Selvarajan V. & K. Rajendran, "Single particle trajectories in a cusped magnetic field"19, 543-545(1991).

[32]Kumar A., V.K. Seneeha & R.M. Vadjikar, " Study of Multi-cusp magnetic field in cylindrical geometry for H - Ion source",AIP Conf. Proc.1097, 137 (2009).

Rashid, M.H., C. Mallik and R.K. Bhandari,
[33]"Simulated new cusp field created by permanent magnet for an 18 GHz ECRIS", Proceedings of the 18th International Conference on Cyclotrons and their applications,pp.286-288.

[34]Schuurman W. & H de Kluiver, " Non-adiabatic motion of a particle in a cusped magnetic field", Journal of Nucl. Energy. Part C,Plasma Phys.7, 245-262, (1964).

[35]Rhee M.J.& W.W. Destler,"Relativistic electron dynamics in a cusped magnetic field",Phys. Fluids 17, 1574 (1974).

[36]Bai,N., Mang Shen, Xiaohan Sun, Fujiang,Liao,and Jinjun "Design of low current density elliptical sheet beam gun", Proceedings of IEEE International Conference on Vacuum Electronics (IVEC),27-29 April 2015 ,Beijing(IEEE,2007),pp.1-2.

[37]Basten,M.A.,Wang Yong and Ding Yaogen,"The Design of Elliptical Sheet Beam Electron Gun",Proceedings of IEEE International Vacuum ElectronicsConference,Kitakyushu,5-7May2007, (IEEE,2007) pp.1-2 .